\documentclass[5p,times]{elsarticle}
\usepackage{hyperref}
\usepackage{graphicx}
\usepackage{amsmath,amssymb}
\usepackage{placeins}
\biboptions{authoryear,round}

\RequirePackage{amsthm,amsmath,amsfonts,amssymb}
\RequirePackage{multirow}
\RequirePackage{booktabs}
\RequirePackage{url}

\theoremstyle{plain}

\theoremstyle{remark}

\begin{document}

\begin{frontmatter}

\title{Nobody Puts Bonferroni in a Corner}

\author{M\aa rten Schultzberg}
\ead{mschultzberg@spotify.com}
\affiliation{organization={Experimentation Platform team, Confidence, Spotify},
  city={Stockholm},
  country={Sweden}}

\begin{abstract}
We argue that Bonferroni correction is a better choice for online experimentation
than it is commonly given credit for. The case rests on four considerations.
First, it is the simplest broadly implementable FWER-controlling method that
produces unconditional simultaneous confidence intervals for every metric.
Second, in a well-specified decision framework, guardrail and quality metrics use
intersection-union logic and cannot inflate the false positive rate, so the
Bonferroni denominator is the number of success metrics only, not the total
metric count. Third, it is uniquely tractable for pre-experiment sample size
calculations. Fourth, we contextualise the power cost empirically. Drawing on a simulation study and an empirical analysis of 1,296 experiments run on Spotify's experimentation platform, Confidence, we show that the power loss relative to more sophisticated FWER methods depends on both how the correction family is specified and how many metrics are truly non-null. When guardrail metrics are incorrectly included in the family, Holm and Hommel are nearly indistinguishable from Bonferroni. When the family is correctly restricted to success metrics only, they gain roughly 4--5 percentage points in ship rate (the fraction of experiments where the treatment is deployed). When few metrics are truly non-null, the gap narrows to near zero regardless of method.
\end{abstract}


\end{frontmatter}

%-----------------------------------------------------------------------
\section{Introduction}
%-----------------------------------------------------------------------

Most experimenters know that testing many hypotheses simultaneously inflates the
false positive rate. Corrections exist to address this, but they carry a
reputation for being costly and inconvenient \citep{perneger1998}, so they are
often skipped. When
they are applied, practitioners sometimes reach for methods that seem more
powerful on paper, without considering whether those methods actually support
the inferences they want to make.

We argue that this reputation is largely undeserved \citep{vanderweele2018},
and that Bonferroni correction specifically is a better choice
for online experimentation than it is commonly given credit for. The organizing claim is simple. Bonferroni is the
simplest broadly implementable FWER-controlling method that remains coherent
across the full experimentation workflow: confidence intervals (CIs), decision
logic, sample size planning, and sequential monitoring. The case rests on four
considerations. First, it is the simplest such method that produces
unconditional simultaneous confidence intervals for every metric. Second, in a
well-specified decision framework, the correction denominator is smaller than it
appears: guardrail and quality metrics use intersection-union logic and cannot
inflate the false positive rate, so the Bonferroni denominator is the number of
\textit{success metrics only}, not the total metric count. Third, it is uniquely
tractable for pre-experiment sample size calculations. Fourth, the power loss
relative to more sophisticated FWER methods depends on both how the correction
family is specified and how many metrics are truly non-null: when guardrail
metrics are incorrectly included in the family, Holm and Hommel are nearly
indistinguishable from Bonferroni; when the family is correctly restricted to
success metrics only, they gain roughly 4--5 percentage points in ship rate (the fraction of experiments
where the treatment is deployed); and when few metrics are truly non-null, the gap
narrows to near zero regardless of
method. Bonferroni also composes cleanly with group sequential testing across
multiple metrics; this is discussed in Appendix~C.

The first three of these (CI coherence, the decision-framework denominator, and
sample size tractability) constitute the paper's core contribution. The fourth
contextualizes the cost.

We support this argument with a simulation study and an empirical analysis of
1,296 experiments (1,840 treatment-control comparisons) run on Spotify's
experimentation platform, Confidence.

%-----------------------------------------------------------------------
\section{Background: The Multiple Testing Problem}
%-----------------------------------------------------------------------

When an experiment tests $m$ hypotheses simultaneously at significance level
$\alpha$, the probability of at least one false positive (the family-wise error
rate, or FWER) inflates well beyond $\alpha$. Under independence,
$\text{FWER} = 1 - (1-\alpha)^m$. For $m = 8$ metrics at $\alpha = 0.05$, this
is approximately 34\%.

A high false positive rate undermines practitioner trust in the experimentation
platform. Practitioners who recognize it begin discounting experiment readouts
and spending time validating results through alternative means, slowing the
platform's influence on decisions. If the elevated rate goes unrecognized,
teams ship ineffective changes, compounding the problem over time.

Several correction methods exist to control FWER or the related false discovery
rate (FDR):

\begin{itemize}
  \item \textbf{Bonferroni}: Test each hypothesis at $\alpha/m$. Requires no
    assumptions about the joint distribution of test statistics; controls FWER
    exactly under independence and conservatively under positive dependence.
  \item \textbf{Holm}: Step-down procedure, uniformly more powerful than
    Bonferroni while still controlling FWER.
  \item \textbf{Hochberg / Hommel}: Step-up procedures, more powerful still,
    with Hommel being the most powerful of the closed-testing family under
    certain conditions.
  \item \textbf{Benjamini--Hochberg (BH)}: Controls FDR rather than FWER. More
    powerful than all of the above, but answers a different question.
  \item \textbf{Benjamini--Yekutieli (BY)}: FDR control without positive
    dependency assumptions. Much more conservative than BH.
\end{itemize}

The narrative in both academic and industry literature tends to cast Bonferroni
as the naive baseline that everyone has moved past, with BH or Hommel positioned
as the modern choice. We argue this framing does not transfer cleanly to the product experimentation
setting, and the FDR versus FWER distinction is where the mismatch is sharpest.

FDR control was developed for genomics, a setting where thousands of hypotheses
are tested simultaneously and many true effects are expected among them. In that
context, permitting a controlled proportion of false discoveries among
significant results is sensible: the goal is discovery at scale, and the false
positives are diluted across a large portfolio. Product experimentation is
structurally the opposite. \citet{berman2022}, analyzing nearly 5,000 effects
from Optimizely experiments, estimate that roughly 70\% of tested effects are
true nulls and that correctly run single-metric tests at 5\% significance
already produce false discovery rates of 18--25\%. This is not a consequence of
statistical malpractice. It is a consequence of the base rate. Most product
changes simply do not move metrics.

In that environment, FDR control's permissiveness compounds the problem rather
than addressing it. FDR methods explicitly allow a higher proportion of false
discoveries among significant results. Applied to a setting where false
discoveries are already endemic due to a high null fraction, they lower a bar
that was already producing too many false positives. A higher shipping rate under
BH is not evidence that more true effects are being captured. It is evidence
that the threshold has been relaxed.

From a decision-theoretic perspective, the choice depends on cost structure and
the prior probability of true effects. Under an additive loss with cost $c_1$
per false positive and $c_2$ per false negative, stringent thresholds are
favoured when $c_1/c_2$ is large, that is, when false positives are
individually costly relative to missed effects. FWER control is designed for
exactly this regime: it bounds the probability of any false positive, which is
appropriate when each false positive causes a specific, consequential error
rather than being diluted across a large portfolio of discoveries. FDR control
was designed for the opposite regime, where many true effects are expected and
some false discoveries among them are tolerable. The high null fraction in
product experimentation undermines the case for FDR on both grounds. When
roughly 70\% of tested effects are null, FDR's permissiveness produces a lower
positive predictive value: a larger share of its additional rejections are false
positives, as a direct consequence of the base rate. A higher ship rate under
BH in this environment is not evidence of recovered true effects; it reflects a
relaxed threshold applied to a setting dominated by nulls.

FWER is generally the right target for product shipping decisions precisely because each
decision is individual and consequential. A false positive is not diluted across
a portfolio of discoveries. It causes a specific team to ship a specific feature
that does nothing, at real engineering and organizational cost.
\citet{kohavi2020} document that trust in the experimentation platform is one of
the most fragile and valuable assets a data-driven organization has, and false
positives erode that trust directly.

%-----------------------------------------------------------------------
\section{Practical Considerations}
%-----------------------------------------------------------------------

\subsection{Confidence Intervals}

A significance test tells you whether to reject. A confidence interval tells you
\textit{what to believe}. In a product experimentation context, the CI is often
more important than the test decision: it communicates effect size, uncertainty,
and practical significance. The CI is what an engineer, product manager, or data
scientist reads when deciding not just \textit{whether} a treatment worked, but
\textit{by how much}.

Bonferroni-corrected confidence intervals are trivial to construct: replace
$\alpha$ with $\alpha/m$ in the standard formula. The resulting intervals are
honest, interpretable, and consistent with the test decision: if a test rejects
at $\alpha/m$, the $1 - \alpha/m$ CI excludes zero, and the $1 - \alpha$ family
of CIs all simultaneously cover their true parameters with probability at least
$1 - \alpha$.

BH does not have this property. Simultaneous confidence intervals for
BH-adjusted tests require a separate and less familiar construction based on the
false coverage rate (FCR) framework of \citet{benjamini2005}. These intervals
are narrower on average, but they are conditional on selection: they cover
parameters among \textit{selected} hypotheses, not all tested hypotheses
simultaneously. Most practitioners are unaware of this distinction, and
experimentation platforms rarely implement FCR-adjusted CIs. In practice,
applying BH and then reporting standard CIs produces intervals that are too
narrow relative to the inferential guarantee actually being provided.

Hommel's procedure is uniformly more powerful than Bonferroni, but its CI story
is limited. Simultaneous CIs via \citet{guilbaud2012} exist only for one-sided
tests, and even then are non-informative for rejected hypotheses: when a metric
reaches significance, the CI merely asserts that the effect lies in the
rejection region rather than providing a numeric bound. For two-sided tests, no
Hommel CIs are available at all. The lack of consistent CIs is not a theoretical footnote: Spotify's
platform, Confidence, only supports Bonferroni
precisely because it is the simplest broadly implementable FWER-controlling
method that delivers coherent CIs universally: for any metric, any direction,
regardless of whether the test is significant. Presenting results in a consistent way is crucial for enabling correct interpretations and building trust in experiment results. \citet{vickerstaff2019} survey the
landscape of multiple comparison methods for RCTs with multiple primary outcomes
and confirm that most methods with higher power do not readily extend to
simultaneous CI construction across all outcomes.

Holm's procedure shares this deficiency \citep{guilbaud2008, strassburger2008}:
the CIs are non-informative precisely for rejected hypotheses, which is when the
winner's curse is a concern. \citet{brannath2024} propose informative CIs that
always yield numeric bounds, but at the cost of modifying the test procedure and
requiring a calibration algorithm that is impractical at platform scale; their
limiting case ($q = 1$) recovers Bonferroni. The European Medicines Agency
guideline, cited by \citet{brannath2024}, explicitly recommends Bonferroni CIs
over the Guilbaud--Strassburger--Bretz construction on these grounds.

Underpowered experiments that reach significance tend to overestimate the true
effect \citep{gelman2014}, and the point estimate alone gives no indication of
how inflated it might be. Although there is no way to avoid the winner's curse, the CI helps with assessing the trustworthiness of the results. Consider two significant results,
both reporting a $+4.5\%$ point estimate. The first has a CI of
$[+4.1\%, +4.9\%]$: the entire range is a positive uplift, and the business
decision is likely the same whether the true effect is at the low or high end. The second has a CI of $[+0.1\%, +9.0\%]$: significant, but spanning
effects from negligible to very large. The first is clearly more trustworthy than the second. A significance indication alone cannot help with this assessment.

\subsection{Decision Framework and the True Correction Denominator}

A more fundamental issue with standard multiple testing discussions is that the
multiple testing problem in product experimentation is often incorrectly framed.
The decision roles of metrics are asymmetric, and that asymmetry changes what
error control actually requires. Correcting for all metrics equally is not
conservative. It addresses the wrong error type for this decision structure.

The typical product experiment has three classes of metrics:

\begin{itemize}
  \item \textbf{Success metrics}: At least one must be statistically significant
    for the experiment to ship. This is a union-intersection (UI) structure.
    Because a false positive on any one success metric could produce a false
    shipping decision, Type~I error control is needed here. Bonferroni divides
    $\alpha$ by the number of success metrics $S$, giving $\alpha^* = \alpha/S$.

  \item \textbf{Guardrail metrics}: All must confirm non-inferiority (where a
    non-inferiority margin is specified) or non-deterioration for the experiment
    to ship. This is an intersection-union (IU) structure. A false positive on a
    guardrail metric (concluding non-inferiority when the metric actually
    degraded) makes it easier to pass that guardrail's gate. But the shipping
    decision requires all guardrails to pass simultaneously. Each additional
    guardrail is an additional constraint, so multiplying guardrails reduces the
    probability of the joint shipping event rather than inflating it. Type~I
    errors on individual guardrail tests cannot increase the false positive rate
    of the overall shipping decision. The multiple testing challenge for
    guardrails runs entirely in the other direction: false negatives (incorrectly
    concluding a metric degraded when it is fine) block good treatments and
    inflate the false negative rate of the overall decision. No Type~I correction
    is needed in the ``shipping direction''; the relevant concern is power, and the appropriate
    adjustment is a beta correction rather than an alpha correction.

  \item \textbf{Quality and deterioration metrics}: These block shipping if they
    move significantly in a negative direction (and include sample-ratio mismatch
    tests, etc.). Like guardrails, their role in
    the decision logic means that false positives lead to over-blocking (shipping
    a good treatment less often), not to false shipping decisions. No Type~I
    correction is needed in the ``shipping direction''.
\end{itemize}

No Type~I correction for guardrails and quality-and-deterioration metrics applies to the shipping decision only. For
the decision to abort an experiment, Type~I correction is needed across all
metrics; see \citet{schultzberg2026} for details.

The guardrail metric argument requires some additional explanation. Let $D$ be the
shipping decision: ship if and only if (i)~at least one success metric is
significant and (ii)~all guardrail metrics confirm non-inferiority. A false
positive on guardrail $g$ means we incorrectly conclude non-inferiority, but
this satisfies condition~(ii) for metric $g$, making it \textit{easier} to ship,
not harder. It therefore contributes to the false positive rate for the overall
decision only if condition~(i) is also satisfied by a true positive on a success
metric. In that case, the treatment genuinely moves a success metric, so
shipping is actually correct. The only remaining path to a problematic false
positive via a guardrail error is a false positive on a \textit{success} metric
combined with a false positive on a guardrail, but the success metric is where
the shipping error originates; the guardrail false positive is irrelevant. Now
consider the symmetric case: a false \textit{negative} on guardrail $g$ means
we incorrectly conclude inferiority when the metric is actually fine. This fails
condition~(ii), blocking a ship even when condition~(i) is met by a genuine
success metric effect. This is precisely the false negative inflation: the
treatment was good, but an overcautious guardrail blocked it. Multiplying
guardrails multiplies this blocking risk, which is a power/beta problem, not a
Type~I problem.

The net result, formalized as Propositions 3.1 and 4.1 in
\citet{schultzberg2026}, is that Bonferroni's denominator for Type~I control is
$S$ (the number of success metrics only), not the total number of metrics in
the experiment. Guardrail and quality metrics contribute to decision quality,
but they do not add to the multiple testing burden on the false positive side.

This matters substantially in practice. Consider an experiment with 2 primary
success metrics and 10 guardrails. Under a naive counting approach, Bonferroni
would divide $\alpha$ by 12. Under a correctly structured decision framework,
the denominator is 2. The correction is $\alpha/2 = 0.025$ rather than
$\alpha/12 \approx 0.004$, a factor of six difference in effective significance
level per metric.

For platforms running large-scale experimentation programs where guardrails
routinely outnumber success metrics, the practical Bonferroni correction is
often very mild. The correction looks severe on paper when the denominator
includes every metric in the experiment; it looks reasonable once the metric
roles are properly accounted for.

Not all platforms use this exact decision structure. What is widespread and unnecessary is including guardrail non-inferiority tests
in the correction family. When a guardrail must pass for the experiment to ship,
false positives on that test make shipping easier, not more likely to be
wrong. Correcting for them inflates the denominator without providing any
Type~I protection on the shipping decision.

Guardrail metrics do affect power, however, and require beta-corrections at the sample size planning stage. See \citet{schultzberg2026} for details.

\subsection{Sample Size Calculations}

High-quality pre-experiment power calculations are essential for running
effective experiment programs. Low-powered experiments are problematic for
several reasons: they risk wasting time and experiment traffic by running
experiments with no plausible chance of reaching a shipping decision, even if the
treatment variant has the intended effect. In addition, the previously mentioned
winner's curse means that even significant experiments inflate the effect
estimates if they have low power, which makes business-impact assessments less
accurate.

Bonferroni is the easiest correction to incorporate in a pre-experiment power
calculation. The adjusted significance level is $\alpha^* = \alpha / S$. In the
simplest case, the required sample size per variant for a two-sample z-test is:

\begin{equation}
  n = \frac{2\sigma^2 (z_{\alpha^*/2} + z_\beta)^2}{\delta^2}
\end{equation}

where $\delta$ is the minimum detectable effect, $\sigma^2$ is the variance, and
$z_{\alpha^*/2}$ is the normal quantile at $\alpha/(2S)$. The correction is a
single substitution. No additional assumptions are needed.

For BH, the picture is more complex. The required sample size depends on the
unknown proportion of true nulls $\pi_0$ and the joint distribution of test
statistics. Under independence, one can derive approximate power formulas
\citep{ferreira2006}, but these require specifying assumptions about the signal
structure that are not available at design time. In practice, BH sample size
calculations are rarely done. Experiments are powered for their primary metric
and the correction is applied post-hoc, with no guarantee about FWER or FDR
properties at the chosen sample size.

This matters. If your platform uses Bonferroni and you pre-specify $S$, you can
make an honest statement: ``This experiment is powered at 80\% to detect a 1\%
relative change in any of $S$ success metrics, controlling FWER at 5\%.'' That
statement is checkable, defensible, and reproducible. The analogous statement
for BH requires assumptions that are almost never verified.

Higher power is of course desirable even if planning is conservative. However,
the willingness to accept the other limitations of these methods lessens
substantially when they cannot help plan more parallel experiments. For companies
like Spotify, where throughput is one of the most important aspects of
experimentation, especially with AI developments, being able to plan and
maximize the number of parallel experiments is critical. 

\subsection{Sequential Testing}

Most industry experimentation platforms monitor experiments over time rather than
running a single analysis at a fixed sample size, which introduces a separate
source of Type~I error inflation through optional stopping. Managing sequential
monitoring and multiple metrics simultaneously is where the choice of correction
method has the most practical consequences. Spotify uses group sequential tests
almost exclusively due to their power advantages over always-valid inference in
batch-analyzed settings \citep{spotify2023}.

The core difficulty under group sequential testing (GST) is that different
metrics do not share a common information fraction at each interim look.
Applying a shared spending function across metrics will over- or underestimate
the information fraction for some, inflating Type~I error or wasting power.
Correlation-exploiting methods such as Hommel face the additional problem that
their sequential boundaries depend on the joint distribution of test statistics
across time, which is rarely available in practice. Bonferroni sidesteps both
issues by allocating an independent $\alpha/m$ budget to each metric, letting
each run its own GST with its own spending function evaluated at its own
information fraction. See Appendix~C for a full discussion, including
always-valid inference alternatives and endpoint-specific information fraction
approaches.

\subsection{Variance Reduction and Metric Correlation}
Variance reduction is used to improve efficiency and can be used in combination
with all multiple testing correction methods. Interestingly, variance reduction
may have a secondary effect on the cross-metric correlation structure, in turn
affecting the effectiveness of the multiple testing correction methods. 

If the
residuals after adjustment are less correlated than the raw metrics (the
condition $\rho_\varepsilon < \rho_0$), then variance reduction reduces
inter-metric correlation as well as per-metric variance, and the practical
conservatism of Bonferroni at $\alpha/S$ is lower than its nominal denominator
implies. This condition is plausible when metric correlation is primarily driven
by stable user-level heterogeneity that variance reduction absorbs, but weakens
when experiment windows are short, metrics are noisy, or metrics have different
temporal dynamics. The decorrelation benefit is therefore strongest in exactly
the conditions where you need it least. The formal derivation and the conditions
under which $\rho_\varepsilon < \rho_0$ holds are given in Appendix~A; the
magnitude of the effect is an empirical question addressed in the supplementary
analysis.

To test this empirically, we re-ran the same experiments through all correction
methods twice: once using each experiment's original variance reduction
configuration (Spotify's implementation follows \citet{negi2021}), and once
forcing plain z-tests for all metrics. Everything else was held constant:
shipping decision logic, significance criterion, system metric exclusion, and
family definition (success metrics only, per Section~3.2).

Bonferroni ship rate: \textbf{23.1\%} with variance reduction, \textbf{21.5\%}
without.

\begin{table*}[t]
\centering
\caption{Ship rates with and without variance reduction. Gap~$\Delta$ = (gap
  vs.\ Bonferroni with variance reduction) $-$ (gap vs.\ Bonferroni without
  variance reduction), in percentage points. Negative indicates variance
  reduction shrank the method's advantage over Bonferroni.}
\label{tab:vr}
\begin{tabular}{lrrrrrr}
\toprule
Method & Ship\% (var.\ red.) & vs.\ Bonf & Ship\% (no var.\ red.) & vs.\ Bonf & Gap $\Delta$ \\
\midrule
BH           & 28.0\% & +4.9pp & 27.2\% & +5.7pp & $-$0.8pp \\
BY           & 27.2\% & +4.1pp & 26.4\% & +4.9pp & $-$0.8pp \\
Hommel       & 27.7\% & +4.6pp & 26.8\% & +5.3pp & $-$0.7pp \\
Holm         & 27.7\% & +4.5pp & 26.8\% & +5.3pp & $-$0.8pp \\
Bonferroni   & 23.1\% & ---    & 21.5\% & ---    & ---      \\
\bottomrule
\end{tabular}
\end{table*}

The decorrelation signal is remarkably uniform: the FDR methods, Holm, and
Hommel all show a gap reduction of 0.7--0.8pp when variance reduction is active (Table~\ref{tab:vr}).
This is consistent with $\rho_\varepsilon < \rho_0$: variance reduction absorbs
shared user-level heterogeneity, modestly reducing the cross-metric correlation
that all alternative methods implicitly exploit relative to Bonferroni. This
pattern is consistent with the theoretical argument in Appendix~A.

\subsection{Actual Power Loss}

The preceding arguments for Bonferroni are partly about what you get for the
cost. But how large is the cost of not using more powerful methods?

\subsubsection{Simulation evidence}

We ran a Monte Carlo simulation with $m = 8$ metrics, $N = 1{,}000$ total
observations, 10,000 replications per effect size, and treatment effects ranging
from 0 to 0.2 standard deviations. Simulation code is available at
\url{https://github.com/MSchultzberg/multiple-testing-paper/}. Tables~\ref{tab:sim0} and~\ref{tab:sim95}
show results under zero and high correlation ($\rho = 0.95$) with all eight
metrics non-null. We additionally vary the number of non-null metrics
$k \in \{1, 4, 8\}$ to assess how the sparsity of true effects changes the
comparison; full results and a block covariance structure appear in
Appendix~B.

Power is defined as the probability of detecting \textit{at least one} true
effect (disjunctive power). All $k$ non-null metrics are assigned the same
effect size $\delta$; heterogeneous effects would activate Holm's step-down
advantage and widen the gap, as seen in the empirical results.

\begin{table*}[t]
\centering
\caption{Simulation results: uncorrelated metrics ($\rho = 0$), all $k = 8$ metrics non-null.}
\label{tab:sim0}
\begin{tabular}{lccccc}
\toprule
Method      & $\delta = 0$ & $\delta = 0.05$ & $\delta = 0.10$ & $\delta = 0.15$ & $\delta = 0.20$ \\
\midrule
None        & 0.340 & 0.835 & 0.992 & 1.000 & 1.000 \\
BH          & 0.052 & 0.326 & 0.830 & 0.997 & 1.000 \\
Hommel      & 0.051 & 0.310 & 0.802 & 0.995 & 1.000 \\
Bonferroni  & 0.051 & 0.303 & 0.783 & 0.991 & 1.000 \\
Holm        & 0.051 & 0.303 & 0.783 & 0.991 & 1.000 \\
BY          & 0.020 & 0.156 & 0.608 & 0.970 & 1.000 \\
\bottomrule
\end{tabular}
\end{table*}

\begin{table*}[t]
\centering
\caption{Simulation results: highly correlated metrics ($\rho = 0.95$), all $k = 8$ metrics non-null.}
\label{tab:sim95}
\begin{tabular}{lccccc}
\toprule
Method      & $\delta = 0$ & $\delta = 0.05$ & $\delta = 0.10$ & $\delta = 0.15$ & $\delta = 0.20$ \\
\midrule
None        & 0.089 & 0.300 & 0.595 & 0.848 & 0.967 \\
BH          & 0.031 & 0.144 & 0.396 & 0.700 & 0.908 \\
Hommel      & 0.026 & 0.131 & 0.368 & 0.676 & 0.895 \\
Bonferroni  & 0.012 & 0.076 & 0.266 & 0.566 & 0.836 \\
Holm        & 0.012 & 0.076 & 0.266 & 0.566 & 0.836 \\
BY          & 0.010 & 0.067 & 0.246 & 0.541 & 0.820 \\
\bottomrule
\end{tabular}
\end{table*}

Results for a block covariance structure (four metrics at $\rho = 0.95$, four
independent) are given in Table~\ref{tab:simblock} in Appendix~B; they sit
between the two extremes shown here.

\subsubsection{Correlation structure and simulation interpretation}

When metrics are correlated, the effective number of independent tests is
smaller than $m$, and the correction is less conservative than its denominator
implies. Correlation-adjusted Bonferroni, for instance via the \citet{nyholt2004}
effective-number formula, is theoretically available as a refinement, but
requires estimating a correlation matrix that varies across experiment
populations and time. The practical recommendation here rests on $\alpha/S$
treating success metrics as independent; the claim is that $S$ is small enough
that this is not materially costly.

This framing also resolves what might seem like a paradox in the simulation
results: all methods lose substantial power under high correlation, and the
methods that nominally exploit correlation (BH, Hommel) do not pull ahead as
much as the zero-correlation gap would suggest. The reason is that under high
correlation there is simply less independent signal across metrics to exploit,
and all methods are working with the same reduced effective information.

\textbf{The gap between BH and Bonferroni is smaller than expected, especially
when metrics are independent.} At a moderate effect size of 0.10 and zero
correlation, BH detects at least one effect in 83.0\% of simulations versus
78.3\% for Bonferroni, a difference of 4.7 percentage points. Hommel sits in
between at 80.2\%. Holm produces identical disjunctive power to Bonferroni
throughout the simulation. This is expected: with equal effect sizes on all
metrics, the binding constraint for disjunctive power is always the minimum
p-value, and both methods apply the same threshold $\alpha/m$ to that minimum.
Holm's step-down advantage only activates when p-values are heterogeneous across
metrics, as in the empirical analysis (Table~\ref{tab:shiprates}), where it ships
4.5pp more than Bonferroni.

\textbf{Under high correlation, the gap widens.} When metrics are highly
correlated, power drops for all methods. BH's disjunctive power advantage over
Bonferroni grows to 13pp at $\delta = 0.10$ (0.396 vs.\ 0.266). Part of what
is happening is structural:
disjunctive power (the probability of detecting \textit{at least one}
effect) mechanically favors more liberal methods, because any single detection
counts as a success. BH's higher power here reflects both its greater liberalism
and its different error target (FDR, not FWER).

\textbf{The gap shrinks further in the sparse regime.} The analyses above assume
all eight metrics are non-null. Table~\ref{tab:prior} shows how the advantage
of Hommel and BH over Bonferroni changes as the number of non-null metrics $k$
varies. When only one metric is truly non-null (the sparse case), both methods
are within 1pp of Bonferroni regardless of correlation structure. The gap opens
as $k$ increases. At $k = 8$ and high correlation, BH pulls 13pp ahead, but under the sparse
conditions that characterize most product experiments, the difference is
negligible. Full results for each $k$ are in Appendix~B.

\begin{table}[t]
\centering
\caption{Disjunctive power advantage over Bonferroni at $\delta = 0.10$, by
  number of non-null metrics and covariance structure.}
\label{tab:prior}
\begin{tabular}{lrrrr}
\toprule
 & \multicolumn{2}{c}{$\rho = 0$} & \multicolumn{2}{c}{$\rho = 0.95$} \\
\cmidrule(lr){2-3}\cmidrule(lr){4-5}
$k$ non-null & Hommel & BH & Hommel & BH \\
\midrule
1 of 8 & $+$0.1pp & $+$0.8pp & $+$0.0pp & $+$0.0pp \\
4 of 8 & $+$1.1pp & $+$3.4pp & $+$2.5pp & $+$6.1pp \\
8 of 8 & $+$2.0pp & $+$4.7pp & $+$10.2pp & $+$13.1pp \\
\bottomrule
\end{tabular}
\end{table}

To be precise about what BH controls under dependence: BH controls FDR under
independence \citep{bh1995} and continues to do so under positive dependence
via the PRDS condition \citep{by2001}. It does not control FWER. The simulation
at $\delta = 0$ with $\rho = 0.95$ shows BH's FWER at 0.038 versus 0.013 for
Bonferroni. Notably, when at most one hypothesis is significant the two error
rates coincide; the gap opens precisely because high correlation makes joint
rejections more likely, inflating FWER beyond what FDR control alone constrains.
In the sparse regime (one true effect out of $m$), a complementary effect
operates: BH's guaranteed FDR $\leq q$ is still satisfied, but the implied FWER
is approximately $2q$; simulations with $m \in \{10, 100, 1000\}$ and a
single non-null at high power give FWER of 8.8--9.6\% against a nominal 5\%
FDR target.
For context, the uncorrected false positive rate under independence is approximately 34\%; Bonferroni brings this to approximately 5\%.

\textbf{The simulation is also not exhaustive.} Results may differ for
non-normal data, heterogeneous effects, small $m$, or very large $m$. We
consider a range of metric counts (2, 10, 50) in the supplementary simulation,
which shows that correlation structure matters more than the raw count of metrics
once $m$ is small enough that corrections are not extreme.

\subsubsection{Empirical evidence from Spotify experiments}

We analyzed the last 1,296 experiments run on Spotify's experimentation
platform, Confidence (1,840 control-treatment comparisons). The dataset
comprises a mix of A/B tests and rollout experiments in the proportion they
naturally occurred on the platform.

A/B tests are designed to answer whether a change should ship: they carry
success metrics that must move significantly, and guardrail metrics that must not
regress. Rollouts are used when a team intends to ship and wants only to verify
that nothing goes wrong. They evaluate guardrail metrics only, and the shipping
decision is whether all non-inferiority tests pass. Including rollouts matters
because the value of any correction method is experienced across the full mix of
experiment types a company runs, not just the subset where statistical power is
the binding constraint.

\textbf{Decision rule.} A comparison ships if: (a)~there are no user-defined
success metrics, or at least one is significant in the preferred direction; and
(b)~there are no guardrail metrics, or all non-inferiority tests pass. Either
condition is vacuously satisfied when the corresponding metric set is
empty (rollouts, which have only guardrail metrics, ship if all NIMs pass).
Significance is defined as adjusted p-value $< \alpha$ throughout. The median
user-defined success metric count per experiment is 2 (mean 3.7) and the median
guardrail metric count is 4 (mean 5.7). Within-experiment metric correlation, estimated from
z-statistics across treatment arms in multi-arm experiments ($n = 398$), is close to zero
(median $\approx 0$, mean $= 0.09$).

\textbf{Shipping rates.} All methods are applied with the success-only family
definition described in Section~3.2: only success metrics enter the correction
family; guardrail metrics are evaluated at their raw p-values and play no role in
inflating the denominator.

\begin{table*}[t]
\centering
\caption{Ship rates by correction method (success-only family, with variance
  reduction). Absolute and relative change versus Bonferroni.}
\label{tab:shiprates}
\begin{tabular}{lrrr}
\toprule
Method        & Ship\% & vs.\ Bonferroni (pp) & vs.\ Bonferroni (rel\%) \\
\midrule
None          & 32.3\% & +9.2pp  & +40\% \\
BH            & 28.0\% & +4.9pp  & +21\% \\
BY            & 27.2\% & +4.1pp  & +18\% \\
Hommel        & 27.7\% & +4.6pp  & +20\% \\
Holm          & 27.7\% & +4.5pp  & +19\% \\
Bonferroni    & 23.1\% & ---     & ---   \\
\bottomrule
\end{tabular}
\end{table*}

The FWER methods no longer cluster at Bonferroni: Holm and Hommel ship at
+4.5--4.6pp above Bonferroni, within a fraction of a point of the FDR methods.
The FDR methods diverge further still, but the gap between FWER and FDR has
narrowed substantially relative to what a naive denominator would suggest. The
FDR methods' higher rate still primarily reflects their different error target
rather than a straightforward power gain: they permit more false discoveries
among significant results by design, and in a setting where the null fraction is
high, the additional ships are consistent with a more permissive error target,
not necessarily with recovery of substantively important true effects.

\subsubsection{Why the family definition is the decisive variable}

To understand what drives the Holm/Hommel result, we ran a crossed comparison:
naive family (all metrics, including guardrails) versus success-only family,
crossed with variance reduction versus no variance reduction.

\begin{table*}[t]
\centering
\caption{Method advantage over Bonferroni (pp) by family definition, with
  variance reduction. Bonferroni: 22.7\% naive, 23.1\% success-only.}
\label{tab:crossed}
\begin{tabular}{lrrr}
\toprule
Method   & Naive family & Success-only family & Gain from correct spec \\
\midrule
BH           & +5.1pp & +4.9pp & $-$0.2pp \\
BY           & +1.8pp & +4.1pp & +2.3pp   \\
Hommel       & +0.6pp & +4.6pp & +4.0pp   \\
Holm         & +0.5pp & +4.5pp & +4.0pp   \\
\bottomrule
\end{tabular}
\end{table*}

Table~\ref{tab:crossed} shows a clear split. Under a naive family, Holm and Hommel are nearly
indistinguishable from Bonferroni (+0.5--0.6pp). Under a success-only family,
they gain +4.5--4.6pp, in line with BH. BH loses slight relative ground: it
gains +5.1pp over Bonferroni under a naive family but only +4.9pp under a
success-only family, consistent with ranking-based methods benefiting more from a
larger p-value list.

The reason comes down to how step-down procedures work. Holm sorts p-values
and progressively relaxes the threshold for each successive test after the
smallest clears its hurdle. When guardrail metrics are included in the family,
they occupy the bottom of the sorted list with large p-values that never
reject, consuming rank positions without contributing to shipping decisions. By
the time the procedure reaches the success metrics, it has spent ranking steps on
tests that were never going to matter. With a success-only family, every rank
position in the step-down corresponds to a metric that can actually drive a ship
decision. The full power of the step-down is available where it is relevant.

This is the empirical counterpart to the theoretical argument in Section~3.2.
The decision framework matters not just because it reduces the denominator for
Bonferroni. It matters because it determines whether FWER-controlling methods
with inherently more structure (step-down, closed testing) can actually exercise
that structure. An experiment with 2 success metrics and 8 guardrails looks,
under a naive correction, like a 10-metric family in which Holm and Bonferroni
are nearly equivalent. Under a correctly specified family, it is a 2-metric
family in which Holm's step-down has full scope to operate, and the power
difference is 4 percentage points.

%-----------------------------------------------------------------------
\section{Why Practitioners Avoid Bonferroni (And Why Those Reasons Are Weaker
Than They Seem)}
%-----------------------------------------------------------------------

The objections to Bonferroni are familiar.

\textbf{``It's too conservative.''} Conservatism is relative to the goal.
Bonferroni controls FWER exactly under independence and is conservative under
positive correlation, where the correction is \textit{less necessary}, not more.

\textbf{``More powerful methods exist.''} True. With a correctly specified
success-only family, Holm and Hommel gain +4.5--4.6pp over Bonferroni in the
empirical analysis. But that gain disappears when guardrail metrics are naively
included in the correction family. Even with a correct family specification
the power gain must be weighted against the fact that neither method produces
simultaneous CIs with numeric bounds for all metrics.

\textbf{``It's not how we do it.''} A platform that has skipped corrections for
years has not been operating cautiously. It has been shipping at an inflated
false positive rate the entire time. Introducing a correction creates
discontinuity: results that would previously have shipped no longer do. That
discontinuity is the correction working.

\textbf{``Bonferroni is genuinely costly when you have many success metrics.''} For teams
testing 8 or more success metrics simultaneously, Bonferroni becomes materially
restrictive: $\alpha/8 = 0.00625$ at the 5\% level, a z-score threshold of
roughly 2.73 rather than 1.96. The argument in this paper weakens in that
regime. But the right response is not to apply Bonferroni mechanically to a
large success set. It is to ask whether that set is well-specified. A collection
of 10 success metrics usually reflects either a lack of prioritization (method
choice cannot fix that) or a set that could be consolidated into a smaller
number of primary metrics plus guardrails. The argument for Bonferroni
is strongest when the success metric set is small and focused, which in our
experience is also the right experimental design practice regardless of which
correction is used.

%-----------------------------------------------------------------------
\section{Conclusion}
%-----------------------------------------------------------------------

This is ultimately an argument about scale and trust. Experiment results do not
sit in a statistical report; they travel through organizations. An effect
estimate is reviewed in a readout meeting, cited in a product decision, compared
to a previous experiment, and referenced in planning documents. At each step, the
inferential framework is summarized once more and potentially simplified once
more. Methods that require careful conditions to produce valid inferences (one-sided
tests only, rejected metrics only, unknown $\pi_0$, shared information
fractions) accumulate approximation errors across these handoffs. Bonferroni
does not. It produces the same honest CI at every step, integrates into sample
size calculations anyone can verify, and pairs with optimal sequential monitoring without
additional assumptions.

Bonferroni is not the most powerful correction available. BH outperforms it on
raw power, and when all metrics are non-null the gap is material under high
correlation. When few metrics are truly non-null — the typical case in product
experimentation — the gap is near zero. The more powerful FWER alternatives (Holm, Hommel) do ship
meaningfully more than Bonferroni when the correction family is correctly
restricted to success metrics: +4.5--4.6pp in the empirical analysis. But that
power gain is conditional on implementing the decision framework in Section~3.2;
without it, Holm and Bonferroni are nearly equivalent in practice. Teams that
do not use that framework forgo the denominator reduction, but not the CI
argument. Bonferroni is the simplest broadly implementable FWER method that
produces honest unconditional simultaneous CIs for every metric, regardless
of how the correction family is defined. For any team that wants to assess whether a result is reliable enough to act
on, that property alone is a strong reason to think twice before choosing a
different method. And even with
a correct family specification, Holm and Hommel cannot match Bonferroni on
unconditional CI coverage or sequential compatibility. Once metric roles are
correctly specified, the denominator is $S$, not the total metric count, which
for most experiments means a mild correction in practice. Each metric runs its
own sequential spending function at its own information fraction. Simple
unconditional simultaneous CIs follow directly. The power gap relative to BH is near zero when few metrics are truly non-null,
and modest even when all metrics move, provided the success metric set is small
and well-defined. That is the right design target regardless of correction
method, and is itself an argument for smaller success sets rather than for
abandoning the correction.

The call is not to avoid corrections. It is to worry less about using Bonferroni.

%-----------------------------------------------------------------------

%-----------------------------------------------------------------------
\appendix
\section{Variance Reduction and Cross-Metric Correlation}
%-----------------------------------------------------------------------

Let post-experiment metrics $A$ and $B$ follow a linear DGP:
\begin{equation}
  A = \mu_A + \gamma_A A_0 + \tau_A T + \varepsilon_A, \quad
  B = \mu_B + \gamma_B B_0 + \tau_B T + \varepsilon_B
\end{equation}
where $A_0, B_0$ are pre-experiment values and $\varepsilon_A, \varepsilon_B$
are residuals uncorrelated with the pre-experiment terms. Variance reduction
yields $\tilde{A} = A - \hat{\gamma}_A A_0$ and $\tilde{B} = B - \hat{\gamma}_B
B_0$, with population residuals $\varepsilon_A, \varepsilon_B$ in large samples.
Under a symmetric parameterization with equal regression coefficients $\gamma$,
equal pre-experiment variance $\sigma_0^2$, and residual variance
$\sigma_\varepsilon^2$, the unadjusted correlation is a weighted average:
\begin{equation}
  \text{Corr}(A, B) = \frac{\gamma^2 \sigma_0^2 \rho_0 + \sigma_\varepsilon^2
  \rho_\varepsilon}{\gamma^2 \sigma_0^2 + \sigma_\varepsilon^2}
\end{equation}
where $\rho_0 = \text{Corr}(A_0, B_0)$ and $\rho_\varepsilon =
\text{Corr}(\varepsilon_A, \varepsilon_B)$. After adjustment,
$\text{Corr}(\tilde{A}, \tilde{B}) = \rho_\varepsilon$. The adjustment therefore
reduces inter-metric correlation if and only if $\rho_\varepsilon < \rho_0$.

The condition $\rho_\varepsilon < \rho_0$ holds when metric correlation is
primarily driven by stable user-level heterogeneity: heavy users have high values
on many metrics, and this shows up in both pre- and post-experiment observations.
Variance reduction removes this shared factor, leaving residuals that reflect
idiosyncratic within-user fluctuations with less reason to co-move. The condition
weakens when correlation is driven by contemporaneous shocks during the
experiment window, when metrics have different temporal dynamics (e.g., a daily
metric and a rolling 7-day metric), or when experiment windows are short and
variance reduction explains little variance, in which case $\rho_\varepsilon
\approx \rho_0$ and the decorrelation benefit is negligible. The decorrelation
benefit from variance reduction is therefore strongest in exactly the conditions
where you need it least: long experiments with stable, high-$R^2$ metrics.

%-----------------------------------------------------------------------
\section{Additional Simulation Tables}
\label{app:sim}
%-----------------------------------------------------------------------
\FloatBarrier

Tables~\ref{tab:simblock}, \ref{tab:prior_k1_indep}, \ref{tab:prior_k1_corr},
\ref{tab:prior_k4_indep}, and~\ref{tab:prior_k4_corr} display additional
simulation results: a block covariance structure and full effect grids for
$k \in \{1, 4\}$ non-null metrics.

\begin{table*}[h!]
\centering
\caption{Simulation results: block-correlated metrics (4 metrics at $\rho = 0.95$,
  4 independent), all $k = 8$ metrics non-null. Results sit between the fully independent and fully correlated cases.}
\label{tab:simblock}
\begin{tabular}{lccccc}
\toprule
Method      & $\delta = 0$ & $\delta = 0.05$ & $\delta = 0.10$ & $\delta = 0.15$ & $\delta = 0.20$ \\
\midrule
None        & 0.252 & 0.694 & 0.966 & 0.999 & 1.000 \\
BH          & 0.045 & 0.251 & 0.719 & 0.978 & 1.000 \\
Hommel      & 0.040 & 0.228 & 0.685 & 0.973 & 1.000 \\
Bonferroni  & 0.037 & 0.216 & 0.653 & 0.961 & 1.000 \\
Holm        & 0.037 & 0.216 & 0.653 & 0.961 & 1.000 \\
BY          & 0.017 & 0.122 & 0.488 & 0.909 & 0.997 \\
\bottomrule
\end{tabular}
\end{table*}

\begin{table*}[h!]
\centering
\caption{Simulation results: $k = 1$ non-null metric, independent ($\rho = 0$).
  Power is P(detecting the one true effect).}
\label{tab:prior_k1_indep}
\begin{tabular}{lccccc}
\toprule
Method      & $\delta = 0$ & $\delta = 0.05$ & $\delta = 0.10$ & $\delta = 0.15$ & $\delta = 0.20$ \\
\midrule
None        & 0.334 & 0.196 & 0.482 & 0.770 & 0.941 \\
BH          & 0.051 & 0.045 & 0.190 & 0.454 & 0.755 \\
Hommel      & 0.050 & 0.043 & 0.184 & 0.446 & 0.751 \\
Bonferroni  & 0.049 & 0.043 & 0.183 & 0.444 & 0.749 \\
Holm        & 0.049 & 0.043 & 0.183 & 0.445 & 0.750 \\
BY          & 0.017 & 0.020 & 0.109 & 0.317 & 0.627 \\
\bottomrule
\end{tabular}
\end{table*}

\begin{table*}[h!]
\centering
\caption{Simulation results: $k = 1$ non-null metric, high correlation ($\rho = 0.95$).
  Power is P(detecting the one true effect).}
\label{tab:prior_k1_corr}
\begin{tabular}{lccccc}
\toprule
Method      & $\delta = 0$ & $\delta = 0.05$ & $\delta = 0.10$ & $\delta = 0.15$ & $\delta = 0.20$ \\
\midrule
None        & 0.088 & 0.200 & 0.468 & 0.766 & 0.935 \\
BH          & 0.030 & 0.049 & 0.176 & 0.451 & 0.743 \\
Hommel      & 0.027 & 0.048 & 0.176 & 0.451 & 0.743 \\
Bonferroni  & 0.013 & 0.045 & 0.176 & 0.451 & 0.743 \\
Holm        & 0.013 & 0.045 & 0.176 & 0.451 & 0.743 \\
BY          & 0.010 & 0.022 & 0.100 & 0.320 & 0.620 \\
\bottomrule
\end{tabular}
\end{table*}

\begin{table*}[h!]
\centering
\caption{Simulation results: $k = 4$ non-null metrics, independent ($\rho = 0$).
  Power is P(detecting at least one of the 4 true effects).}
\label{tab:prior_k4_indep}
\begin{tabular}{lccccc}
\toprule
Method      & $\delta = 0$ & $\delta = 0.05$ & $\delta = 0.10$ & $\delta = 0.15$ & $\delta = 0.20$ \\
\midrule
None        & 0.330 & 0.581 & 0.925 & 0.997 & 1.000 \\
BH          & 0.051 & 0.171 & 0.587 & 0.930 & 0.997 \\
Hommel      & 0.051 & 0.163 & 0.564 & 0.919 & 0.995 \\
Bonferroni  & 0.050 & 0.161 & 0.554 & 0.912 & 0.995 \\
Holm        & 0.050 & 0.161 & 0.555 & 0.912 & 0.995 \\
BY          & 0.018 & 0.082 & 0.372 & 0.816 & 0.984 \\
\bottomrule
\end{tabular}
\end{table*}

\begin{table*}[h!]
\centering
\caption{Simulation results: $k = 4$ non-null metrics, high correlation ($\rho = 0.95$).
  Power is P(detecting at least one of the 4 true effects).}
\label{tab:prior_k4_corr}
\begin{tabular}{lccccc}
\toprule
Method      & $\delta = 0$ & $\delta = 0.05$ & $\delta = 0.10$ & $\delta = 0.15$ & $\delta = 0.20$ \\
\midrule
None        & 0.087 & 0.271 & 0.558 & 0.836 & 0.960 \\
BH          & 0.030 & 0.092 & 0.290 & 0.623 & 0.863 \\
Hommel      & 0.026 & 0.074 & 0.254 & 0.579 & 0.838 \\
Bonferroni  & 0.014 & 0.064 & 0.229 & 0.547 & 0.817 \\
Holm        & 0.014 & 0.064 & 0.229 & 0.547 & 0.817 \\
BY          & 0.010 & 0.042 & 0.177 & 0.466 & 0.761 \\
\bottomrule
\end{tabular}
\end{table*}

\FloatBarrier

%-----------------------------------------------------------------------
\section{Sequential Testing and Group Sequential Tests}
\label{app:seq}
%-----------------------------------------------------------------------

Most industry experimentation platforms monitor experiments over time rather
than running a single analysis at a fixed sample size. Three main approaches
exist for handling the sequential dimension: always-valid inference (AVI), group
sequential tests (GST), and Bonferroni-over-time. Spotify uses GST almost
exclusively due to its power advantages over AVI in batch-analyzed settings
\citep{spotify2023}; this appendix focuses on that setting.

\subsection*{Always-valid inference}

Always-valid (or anytime-valid) methods, such as the mSPRT of \citet{johari2022},
produce p-values that are valid at any stopping time. Combining them with a
multiple testing correction is straightforward: apply Bonferroni across metrics
to anytime-valid p-values and validity is preserved at all times.

The cost is power. AVI methods are built for continuous monitoring and are
conservative when data is analyzed in batches rather than after every
observation. Spotify's analysis of sequential testing frameworks found that AVI
methods achieve power of roughly 0.72--0.76 compared to approximately 0.90 for
GST under a typical effect size and sample size \citep{spotify2023}. Even a
simple Bonferroni-over-time approach with 14 intermittent analyses outperforms
AVI methods in batch settings.

\subsection*{Group sequential tests and the information fraction problem}

GST methods based on the Lan--DeMets alpha spending function \citep{lan1983} are
the most powerful sequential framework for batch-analyzed experiments. The
problem with multiple metrics under GST is fundamental: \textbf{different
metrics do not share a common information fraction at each interim analysis.}

The information fraction at a given calendar time is metric-specific. Using the
primary metric's information fraction for all metrics will overestimate it for
some and underestimate it for others: overestimation inflates Type~I error;
underestimation costs power. The problem compounds when metrics include both
ratio metrics (like revenue per user) and proportion metrics (like conversion
rate), which can have quite different effective sample sizes at any given point.

A concrete example: consider an experiment running for three weeks, measuring
one-day activity (whether a user was active in their first day of exposure) and
second-week consumption (total streams in days 8--14). After four days, one-day
activity has been observed for nearly every user (information fraction
approaching 90\% or higher), while second-week consumption is exactly zero; no
user has yet entered their second week. By day 14, the second metric's
information fraction is still roughly 50\% while the first is close to 1. These
two metrics cannot share a spending function.

Solutions exist for specific settings. \citet{liao2018} propose defining
endpoint-specific information fractions: separate spending functions for each
endpoint, each evaluated at its own information fraction. \citet{gao2008} extend
this to adaptive re-estimation. Both approaches require upfront specification of
variance structures and data maturity timelines across all metrics, which is
rarely feasible in product experimentation.

Correlation-exploiting methods like Hommel face a further obstacle: their
sequential boundaries depend on the correlation structure of the test statistics
\textit{across time}, not just at the final analysis, and this cannot be
specified without joint distributional assumptions that are rarely available.

\subsection*{Bonferroni's tractability}

Bonferroni sidesteps these problems by allocating an independent $\alpha/m$
budget to each metric, letting each run a separate GST with its own alpha
spending function evaluated at its own information fraction. No shared
information fraction is required, no cross-metric dependence structure needs to
be specified, and each metric's sequential analysis is fully decoupled from the
others. Run each metric's sequential test at level $\alpha/m$ using an
O'Brien--Fleming or Pocock-type boundary evaluated at that metric's own
information fraction at each look. The resulting procedure controls FWER across
metrics and time simultaneously, with no cross-metric dependencies to manage.

\end{document}